\pdfoutput=1
\documentclass[10pt]{article}
\usepackage{graphicx}

\def\Title#1{\begin{center} {\Large #1 } \end{center}}
\def\Author#1{\begin{center}{ \sc #1} \end{center}}
\def\Address#1{\begin{center}{ \it #1} \end{center}}

\newcommand\pubblock{\rightline{\begin{tabular}{l} Proceedings of the Second Annual LHCP\\ \pubnumber\\
         \pubdate  \end{tabular}}}

\newenvironment{Abstract}{\begin{quotation} \begin{center} 
             \large ABSTRACT \end{center}\bigskip 
      \begin{center}\begin{large}}{\end{large}\end{center} \end{quotation}}

\newenvironment{Presented}{\begin{quotation} \begin{center} 
             PRESENTED AT\end{center}\bigskip 
      \begin{center}\begin{large}}{\end{large}\end{center} \end{quotation}}

\def\beq{\begin{equation}}
\def\eeq#1{\label{#1}\end{equation}}
\def\eeqn{\end{equation}}

\def\beqa{\begin{eqnarray}}
\def\eeqa#1{\label{#1}\end{eqnarray}}
\def\eeqan{\end{eqnarray}}

\let\bar=\overbar

\def\Dslash{\not{\hbox{\kern-4pt $D$}}}
\def\dslash{\not{\hbox{\kern-2pt $\del$}}}

\def\mh{m_h}

\def\msb{{\bar{\ssstyle M \kern -1pt S}}}

\def\pt{\ensuremath{p_{\mathrm{T}}}}
\def\ptmiss{\ensuremath{\pt^{\mathrm{miss}}}}
\def\Mmumu{\ensuremath{m_{\mu\mu}}}
\def\mm{\ensuremath{\mu^+\mu^-}}
\def\hmm{\ensuremath{\mathrm{H}\to\mu^+\mu^-}}
\def\mh{\ensuremath{\mathrm{m_H}}}
\def\TeV{\ensuremath{\mathrm{Te\kern -0.1em V}}}
\def\GeV{\ensuremath{\mathrm{Ge\kern -0.1em V}}}
\def\GeVc{\GeV{}}
\def\GeVcc{\GeV{}}

\usepackage{color}

\newcommand\pubnumber{ CMS CR-2014/189 }

\newcommand\pubdate{\today}

\def\affiliation{
On behalf of the CMS Experiment, \\
Department of Physics \\
University of Florida, Gainesville, FL, USA }

\begin{document}

\large
\begin{titlepage}
\pubblock

\vfill
\Title{ Search for the Standard Model Higgs Boson Decaying to $\mu^+\mu^-$ in $pp$ Collisions at $\sqrt{s}=7$ and 8\,\TeV{} with the CMS Detector }
\vfill

\Author{ Justin Hugon  }
\Address{\affiliation}
\vfill
\begin{Abstract}

A search for the standard model Higgs boson in the rare $\mu^+\mu^-$ decay channel is presented. The data samples, recorded by the CMS experiment at the LHC, correspond to integrated luminosities of $5.0\pm0.1$\,fb$^{-1}$ at 7\,\TeV{} center-of-mass energy and of $19.7\pm0.5$\,fb$^{-1}$ at 8\,\TeV{}. To enhance the Higgs signal over the dominant Drell-Yan background, the events are categorized by topologies corresponding to different production processes. Upper limits on the production rate, with respect to the Standard Model prediction, are reported at the 95\% confidence level for Higgs boson masses in the range from 120 to 150\,\GeVcc{}.

\end{Abstract}
\vfill

\begin{Presented}
The Second Annual Conference\\
 on Large Hadron Collider Physics \\
Columbia University, New York, U.S.A \\ 
June 2-7, 2014
\end{Presented}
\vfill
\end{titlepage}
\def\thefootnote{\fnsymbol{footnote}}
\setcounter{footnote}{0}
%

\normalsize 


\section{Introduction}

The recent discovery of a particle with properties consistent with a 
standard model (SM) Higgs boson~\cite{Aad:2012tfa,Chatrchyan:2012ufa} 
has provided new insight into the nature of the SM and possible new physics extensions.
More may be learned by further study of the production and decay modes of
the new particle, especially highly suppressed ones.  The SM Higgs boson decay
to a pair of opposite sign muons is a perfect candidate, due to the small
SM branching ratio ($2.2\times10^{-4}$~\cite{Denner:2011mq}), combined
with the clean final state.  

This document reports on a search for a SM-like Higgs boson in the \mm{} decay channel 
with the CMS experiment at the CERN LHC.  
This search has previously been reported in more detail in Ref.~\cite{CMS:2013aga}.
Data samples corresponding to integrated luminosities of 
$5.0\pm0.1$\,fb$^{-1}$ at a center-of-mass energy ($\sqrt{s}$) of
7\,\TeV{} and $19.7\pm0.5$\,fb$^{-1}$ at 8\,\TeV{} are analyzed.
The Higgs boson signal is sought as a narrow signal peak in the 
dimuon invariant mass (\Mmumu{}) spectrum, over a smoothly falling
background dominated by the Drell--Yan and $t\bar{t}$ production processes.  
An event categorization is used to increase search sensitivity.  
The signal efficiency is estimated using a Monte-Carlo
simulation which has been extensively validated on data.
The rates of signal and background production are estimated by fitting 
parameterized shapes to the \Mmumu{} spectra in each category.

\section{Search Methodology}

Events are collected with a trigger that requires at least one
isolated muon candidate with transverse-momentum (\pt) greater than 24\,\GeVc{},
and $|\eta|<2.1$, where $\eta$ is the pseudorapidity.  In the offline
data analysis, events are required to contain two opposite-sign
muon candidates, each with $|\eta|<2.1$.  The muon candidate that passed 
the trigger selection must have $\pt{}>25$\,\GeVc{}, while the other must 
have $\pt>15$\,\GeVc{}. Both muon candidates must also pass isolation and quality 
criteria.

Events passing the above selection are then assigned to categories.  Events
where the leading jet \pt{} is greater than 50\,\GeVc{}, the sub-leading
jet \pt{} is greater than 30\,\GeVc{}, and the \ptmiss{} is less than 40\,\GeVc{} 
are assigned to the 2-Jet category.  The \ptmiss{} is defined as the transverse
momentum of the sum of the dimuon and dijet four-vectors.
All events not assigned to the 2-Jet category are assigned to the 0,1-jet category.
The 2-Jet category has a high signal-to-background ratio (S/B) due to the 2-Jet
topology of VBF Higgs boson production, while the 0,1-jet category has a large
efficiency for GF Higgs boson production.

Events in the 2-Jet category are further assigned to one of three subcategories.  The 2-Jet
VBF Tight subcategory is optimized for VBF production, the 2-Jet GF Tight
category is optimized for GF production, while the remaining 2-Jet events
are assigned to the 2-Jet Loose category.  Events in the 0,1-Jet categories
are assigned to one of 15 subcategories depending on the dimuon \pt{} and the $|\eta|$
of the muons.  The dimuon \pt{} helps in separating the GF Higgs boson signal from background,
while muon $|\eta|$ is related to the \Mmumu{} resolution.  The \Mmumu{}
mass resolution for different categories varies from 3.8 to 5.9\,\GeVcc{}, 
in terms of the signal peak full-width-at-half-maximum at $\mh=125\,\GeVcc{}$.

\begin{figure}[htb]
\centering
\includegraphics[width=0.38\textwidth]{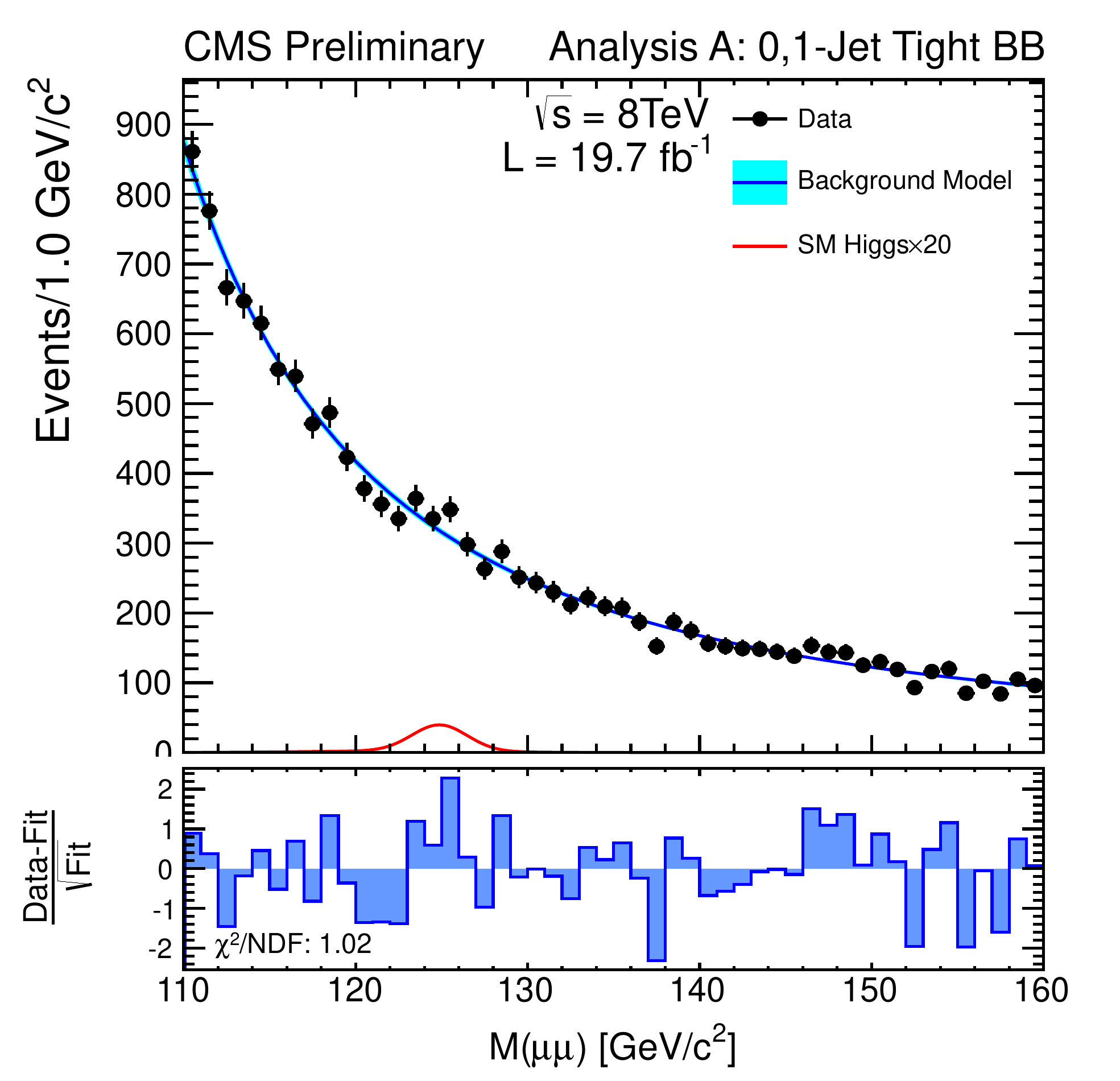}
\includegraphics[width=0.38\textwidth]{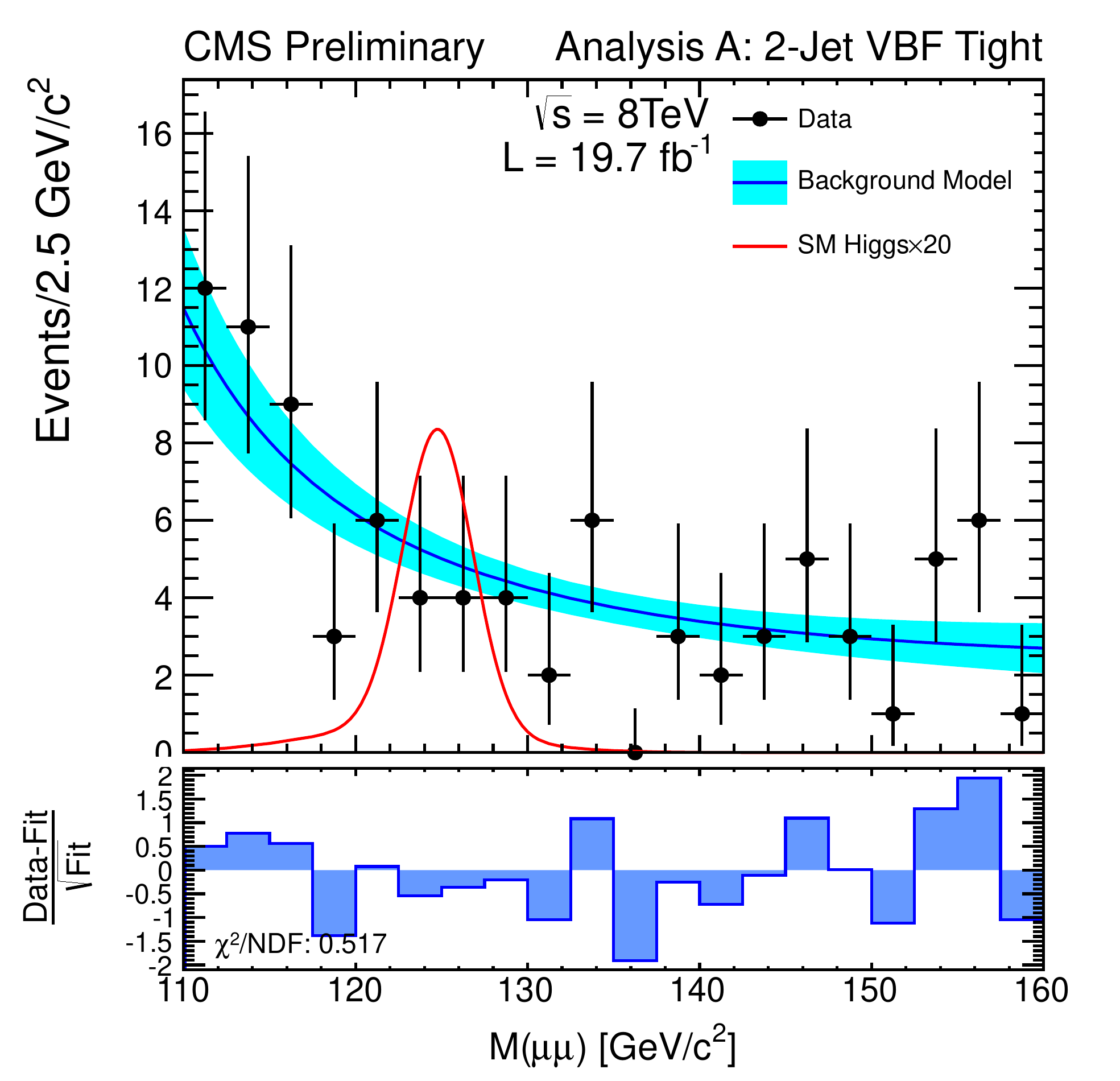}
\caption{ The dimuon invariant mass distribution for the 0,1-Jet Tight BB (left) and 2-Jet VBF Tight (right) categories.
            A fit of the parameterized background model used to estimate the background contribution is shown
            as a dark blue line, with its uncertainty shown as a light blue band.  The expected SM Higgs boson
            signal magnified by a factor of twenty is shown by the red line, for $\mh{}=125\,\GeVcc{}$.
            }
\label{fig:mass}
\end{figure}

The rates of signal and background production are then extracted from the 
\Mmumu{} distribution 
in each category.  A parameterized signal shape made of the sum of
two Gaussian functions is used to model the shape of the Higgs boson signal, while
the function,
\begin{equation}
f(\Mmumu{}) = \frac{\mathrm{e}^{\,p_1 \Mmumu{}}}{\Mmumu{}-p_2},
\end{equation}
is used to model the shape of the background.
Example fits of the background model to the data are shown in Figure~\ref{fig:mass}.

\section{Results}

\begin{figure}[htb]
\centering
\includegraphics[width=0.38\textwidth]{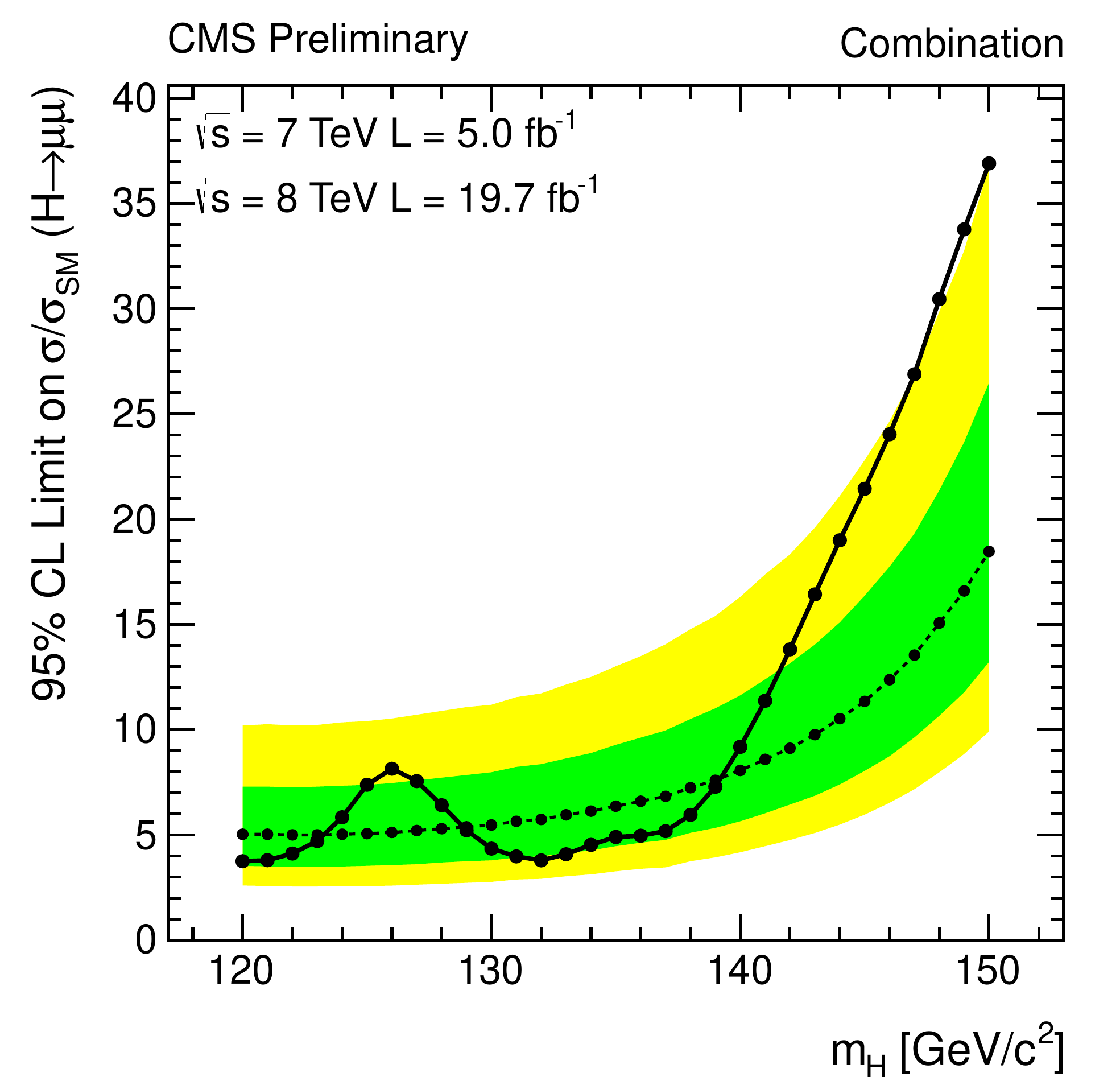}
\includegraphics[width=0.38\textwidth]{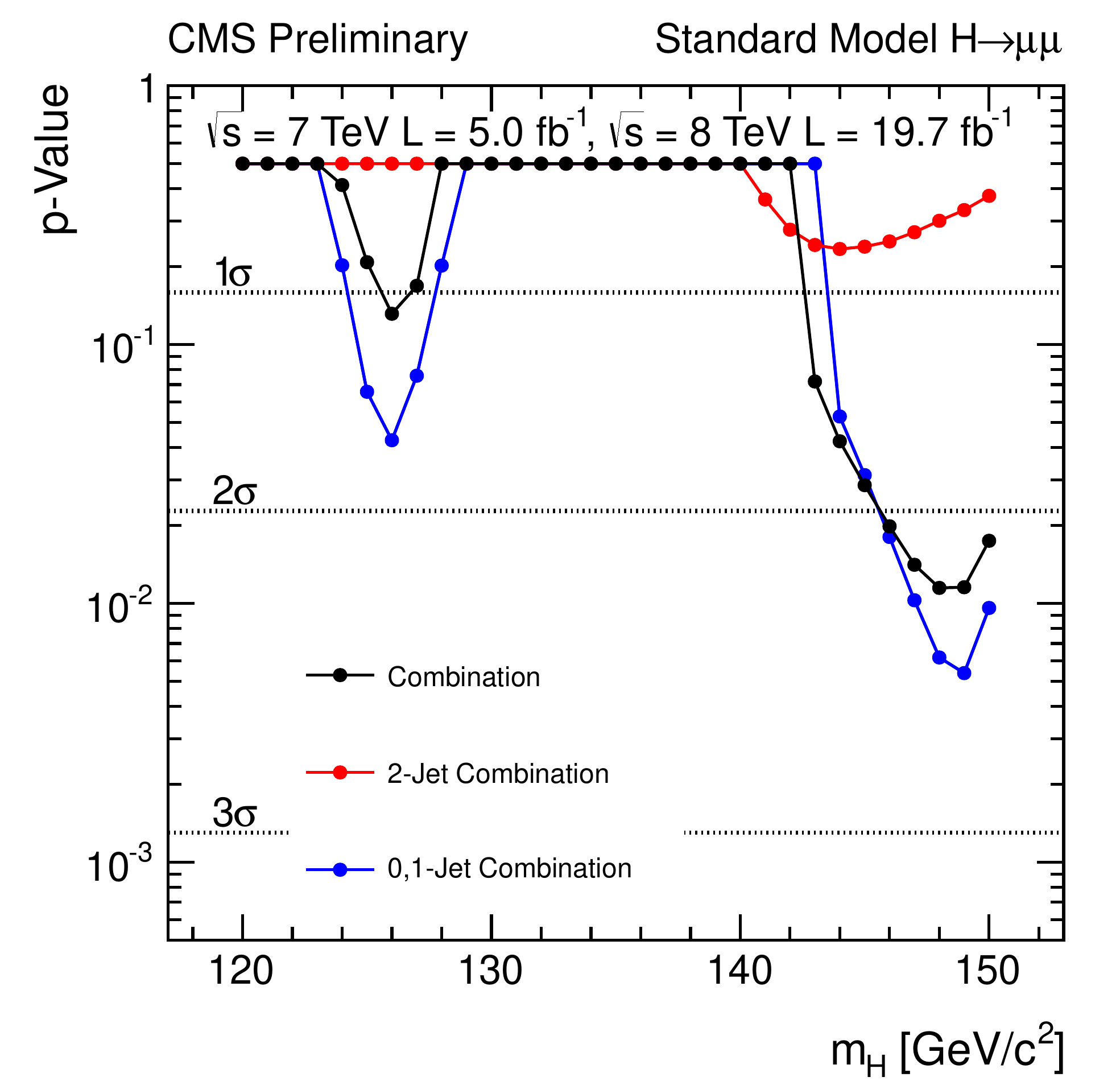}
\caption{ Upper limits on the signal strength (left) and p-values for the 
            background-only hypothesis (right) for \mh{} from 120 to 150\,\GeVcc{}.}
\label{fig:results}
\end{figure}

Results are extracted by simultaneous fits of the parameterized signal and background shapes
to the \Mmumu{} distributions in each category.  Upper limits
on the rate of \hmm{} production divided by the expected SM \hmm{} rate ($\sigma/\sigma_\mathrm{SM}$) 
for a Higgs boson mass between 120 and 150\,\GeVcc{} are shown in Figure~\ref{fig:results} at 
the 95\% confidence level.  
For $\mh{}=125\,\GeVcc{}$, an expected (observed) 95\% confidence level upper limit
on the rate of \hmm{} production is set at 
$5.1^{+2.3}_{-1.5}\times\mathrm{SM}$ ($7.4\times\mathrm{SM}$).
For a Higgs boson mass between 120 and 150\,\GeVcc{}, background-only p-values are shown
in Figure~\ref{fig:results}.  A small excess of events is observed near $\mh{}=125\,\GeVcc{}$, with
a significance of about $1\sigma$.  A larger excess is observed around 148\,\GeVcc{}
with a local significance of $2.3\sigma$ and a global significance
of $0.8\sigma$.

Assuming a SM-like Higgs boson of mass 125.7\,\GeVcc{} decays to \mm{},
the best fit signal rate is $2.9^{+2.8}_{-2.7}\times\mathrm{SM}$.

\section{Projections to $\mathbf{\sqrt{s}=14}$\,\TeV{}}

This section presents the results of Ref.~\cite{CMS:2013xfa} in relation to
the projected performance of the CMS \hmm{} search at $\sqrt{s}=14$\,\TeV{}.
The methodology of the projections is to use the current \hmm{} search
procedure and simulated data samples without modification, except
for scaling the signal and background cross sections from $\sqrt{s}=8$\,\TeV{}
to 14\,\TeV{} and similarly scaling the luminosity.
In addition, two systematic uncertainty scenarios are investigated.
In Scenario~1, all systematic uncertainties are assumed to be the same
as in the current search, while in Scenario~2, the experimental uncertainties
are scaled by $1/\sqrt{\mathcal{L}}$, where $\mathcal{L}$ is the integrated luminosity, 
while the theoretical uncertainties are reduced by 50\%.

The results of the projections are that SM \hmm{} can be excluded with 
$175\pm^{+150}_{-75}$\,fb$^{-1}$ of integrated luminosity, for $\mh{}=125\,\GeVcc{}$.  
The two scenarios differ little at this luminosity.  
Projections for 300 and 3000\,fb$^{-1}$ are shown in Table~\ref{tab:projTab}.  
The table shows, for $\mh{}=125\,\GeVcc{}$, the expected significance of SM \hmm{}, the uncertainty
on the best fit of the \hmm{} rate, and the uncertainty on the best fit
of the coupling of muons to the Higgs boson, when combined with other CMS Higgs boson measurements.

\begin{table}[t]
\begin{center}
\caption{Projected \hmm search sensitivity at $\sqrt{s}=14$\,\TeV{} for integrated 
            luminosities of 300 and 3000\,fb$^{-1}$.  See text for details.}
\label{tab:projTab}
  \begin{tabular}{lcc} \hline
      & 300\,fb$^{-1}$ & 3000\,fb$^{-1}$ \\ \hline
    Significance  & $2.5\sigma$ & $7.9\sigma$ \\ 
    Uncertainty on $\sigma/\sigma_{SM}$ Scenario 1  & 42\% & 20\% \\ 
    Uncertainty on $\sigma/\sigma_{SM}$ Scenario 2  & 40\% & 14\% \\ 
    Uncertainty on Muon Coupling Scenario 1  & 23\% & 8\% \\ 
    Uncertainty on Muon Coupling Scenario 2  & 23\% & 7.5\% \\ \hline
  \end{tabular}
\end{center}
\end{table}

\section{Summary}

A search for a SM-like Higgs boson in the \mm{} decay channel is reported.
A narrow \hmm{} peak in the \Mmumu{} spectrum is sought over a smoothly
falling background dominated by Drell--Yan and $t\bar{t}$.  Search
sensitivity is enhanced with an event categorization.  The
signal and background rates are extracted by fitting parameterized
signal and background shapes to the \Mmumu{} spectra in each category.

No significant excess is found, so upper limits are set on the
rate of \hmm{} production with respect to the SM prediction.
For $\mh{}=125\,\GeVcc{}$, a 95\% confidence level upper limit
on \hmm{} production is set at $7.4\times\mathrm{SM}$.
Assuming a SM-like Higgs boson of mass 125.7\,\GeVcc{} decays to \mm{},
the best fit rate is $2.9^{+2.8}_{-2.7}\times\mathrm{SM}$.

Projections to $\sqrt{s}=14$\,\TeV{} are also reported.  For $\mh{}=125\,\GeVcc{}$,
SM Higgs boson decays to \mm{} are projected to be excluded with
$175\pm^{+150}_{-75}$\,fb$^{-1}$ of integrated luminosity.  The expected significance
of the SM \hmm{} signal is projected to be 
$2.5\sigma$ ($7.9\sigma$) for 300\,fb$^{-1}$ (3000\,fb$^{-1}$) of integrated luminosity.



\begin{thebibliography}{9}


\bibitem{Aad:2012tfa} 
  G.~Aad {\it et al.}  [ATLAS Collaboration],
  ``Observation of a new particle in the search for the Standard Model Higgs boson with the ATLAS detector at the LHC,''
  Phys.\ Lett.\ B {\bf 716}, 1 (2012)
  [arXiv:1207.7214 [hep-ex]].
  
\bibitem{Chatrchyan:2012ufa} 
  S.~Chatrchyan {\it et al.}  [CMS Collaboration],
  ``Observation of a new boson at a mass of 125\,\GeV{} with the CMS experiment at the LHC,''
  Phys.\ Lett.\ B {\bf 716}, 30 (2012)
  [arXiv:1207.7235 [hep-ex]].

\bibitem{Denner:2011mq} 
  A.~Denner {\it et al.},
  ``Standard Model Higgs-Boson Branching Ratios with Uncertainties,''
  Eur.\ Phys.\ J.\ C {\bf 71}, 1753 (2011)
  [arXiv:1107.5909 [hep-ph]].


\bibitem{CMS:2013aga} 
  CMS Collaboration,
  ``Search for the standard model Higgs boson in the $\mu^+\mu^-$ decay channel 
in $pp$ collisions at $\sqrt{s}= 7$ and 8\,\TeV{},''
  CMS-PAS-HIG-13-007.

\bibitem{CMS:2013xfa} 
  CMS Collaboration,
  ``Projected Performance of an Upgraded CMS Detector at the LHC and HL-LHC: Contribution to the Snowmass Process,''
  [arXiv:1307.7135 [hep-ex]].

\end{thebibliography}
\end{document}